\documentclass{article}

\usepackage[a4paper,top=3cm,bottom=2cm,left=2cm,right=2cm,marginparwidth=1.75cm]{geometry}

\newcommand{\Ccal}{\mathcal{C}}

\newcommand{\Hcal}{\mathcal{H}}

\newcommand{\red}[1]{\textcolor{red}{#1}}
\newcommand{\blue}[1]{\textcolor{blue}{#1}}
\newcommand{\green}[1]{\textcolor{darkgreen}{#1}}
\newcommand{\TMH}[2]{{#2}}

\newcommand{\SR}[2]{{#2}}

\usepackage{xcolor}
\usepackage{amsfonts}
\usepackage{latexsym}
\usepackage{amsmath, amssymb}
\usepackage{graphicx}
\usepackage{color}
\usepackage{natbib}
\usepackage{booktabs}
\usepackage{colortbl}
\usepackage{url}
\usepackage{ulem}
\definecolor{darkgreen}{rgb}{0,0.55,0}

\title{Querying multiple sets of $p$-values through composed hypothesis testing}

\author{Tristan Mary-Huard$^{1,2}$, Sarmistha Das$^{3}$, \\
Indranil Mukhopadhyay$^{3}$ and St\'ephane Robin$^{1,4}$}

\date{}

\begin{document}

\maketitle

\noindent$(^{1})$ MIA-Paris, INRAE, AgroParisTech, Universit\'e Paris-Saclay, Paris, 75005, France. \\
$(^{2})$ GQE-Le Moulon, Universite Paris-Saclay, INRAE, CNRS, AgroParisTech, Gif-sur-Yvette, 91190, France.\\
$(^{3})$ Human Genetics Unit, Indian Statistical Institute, Kolkata, 700108, India. \\
$(^{4})$ Centre d'\'Ecologie et des Sciences de la Conservation (CESCO), MNHN, CNRS, Sorbonne Universit\'e, Paris, 75005, France. \\

\abstract{
\noindent \textbf{Motivation:} 
Combining the results of different experiments to exhibit complex patterns or to improve statistical power is a typical aim of data integration. The starting point of the statistical analysis often comes as sets of $p$-values resulting from previous analyses, that need to be combined in a flexible way to explore complex hypotheses, while guaranteeing a low proportion of false discoveries. \\
\textbf{Results:} 
We introduce the generic concept of {\sl composed hypothesis}, which corresponds to an arbitrary complex combination of simple hypotheses. We rephrase the problem of testing a composed hypothesis as a classification task, and show that finding items for which the composed null hypothesis is rejected boils down to fitting a mixture model and classify the items according to their posterior probabilities. We show that inference can be efficiently performed and provide a thorough classification rule to control for type I error. The performance and the usefulness of the approach are illustrated on simulations and on two different applications combining data from different types. The method is scalable, does not require any parameter tuning, and provided valuable biological insight on the considered application cases. 

\paragraph{Availability:} We implement the QCH methodology in the \texttt{qch} R package hosted on CRAN. 
}

\paragraph{Contact:} \url{tristan.mary-huard@agroparistech.fr} 


\section{Introduction \label{sec:Intro}} 
\paragraph{Combining analyses.}
Since the beginning of omics era it has been a common practice to compare and intersect lists of $p$-values, where all lists describe the same items (say genes) whose differential activity was tested and compared in different conditions or using different omics technologies. One may e.g. consider $i$) genes whose expression has been measured in $Q$ different types of tissues (and compared to a reference), $ii$) genes whose expression has been measured in a same tissue at $Q$ different timepoints (compared to baseline), or $iii$) genes whose activity has been investigated through $Q$ omics such as expression, methylation and copy number. The goal of the post-analysis is then to identify items that have been perturbed in either all or \SR{only some selected}{in a predefined subset of} conditions. Finding such genes by integrating information from multiple data sources may be a first step towards understanding the underlying process at stake in the response to treatment, or the inference of the undergoing regulation network  \cite{Das2019,Xiong2012}.

\paragraph{List crossing.}
\SR{The simplest}{One simple} way to cross sets of $p$-values is to simply apply a multiple testing procedure separately to each list, identify the subsets of items for which the $H_0$ hypothesis is rejected, then intersect these subsets. The graphical counterpart of this strategy is the popular Venn diagram that can be found in numerous articles (see e.g. \cite{Conway2017} and references inside). Alternatively, a lot of attention has been dedicated to the consensus ranking problem, where one aims at combining multiple ranked lists into a single one that is expected to be more reliable, see \cite{Li2019} for an extensive review. 
\SR{However the question one aims to answer through these intersecting or consensus ranking approaches is often unclear.}{However, neither intersecting nor consensus ranking rely on an explicit biological hypothesis.}
Moreover, apart from rare exceptions such as \cite{Natarajan2012}, in most cases the final set of identified items comes with no statistical guarantee regarding false positive control. 

\paragraph{Composed hypotheses.}
The primary objective of the present article is to provide a statistical framework suitable to answer complex queries called hereafter {\sl composed hypotheses}, such as "which genes are expressed in a subset of conditions ?". This corresponds to a complex query because it cannot be answered through standard testing procedures. A classical example of a composed hypothesis is the so called Intersection-Union Test (IUT) \citep{Berger1996} where one aims at finding the items for which all the $Q$ tested hypotheses should be rejected, e.g. genes declared differentially expressed for all treatment comparisons. \\
\SR{This}{Testing composed hypotheses on a large set of items} requires $i$) the identification of a proper test statistic to rank the items based on their $p$-value profile and $ii$) a thresholding rule that provides guarantees about type I error rate control. Although different methods have been proposed to build the test statistic and to control the type I error rate in the specific context of the IUT problem (\cite{Deng2008,Tuke2009,Van2009}), no generic framework has emerged to easily handle any arbitrary composed hypothesis so far.    

\paragraph{Contribution.}
We propose to perform composed hypothesis testing using a mixture model, where each item belongs to a class characterized by a combination of $H_0$ and $H_1$. The strategy we propose is efficient in many ways. First, it comes with a natural way to rank the items and control type I error rate through their posterior probabilities and their local \SR{FDR}{False Discovery Rate (FDR)} interpretation. Second, we show that, under mild conditions on the conditional distributions, inference can be efficiently performed on a large collection of items (up to several millions in our simulations) within seconds, making the method amenable to applications such as meta-analysis of \SR{GWA}{genome-wide association} studies. 
The method consists in three independent steps: first fit a non-parametric mixture model on each marginal distribution, then estimate the proportion of the joint mixture model and finally query the composed hypothesis. Importantly, the first two fitting steps are performed only once and can then be used to answer any number of composed hypothesis queries without additional computational burden. 
Lastly, using both simulated and real genomic data, we illustrate the performance of the strategy (in terms of FDR control and detection power) and its richness in terms of application. In particular for the classical IUT problem it is shown that the detection power improves with respect to the number of $p$-value sets up to almost 100\% in some cases where a classical list crossing would be almost inefficient.  

\section{Model \label{sec:Model}} 
\paragraph{Composed hypothesis.} 
We consider the test of a composed hypothesis relying on $Q$ different tests. More specifically, we denote by $H^q_0$ (resp. $H^q_1$) the null (resp. alternative) hypothesis corresponding to test $q$ ($1 \leq q \leq Q$) and consider the set $\Ccal := \{0, 1\}^Q$ of all possible combinations of null and alternative hypotheses across the $Q$. We name {\sl configuration} any element $c := (c_1, \dots c_Q) \in \Ccal$. There exist $|\Ccal| = 2^Q$ such configurations. For a given configuration $c$, we define the {\sl joint hypothesis} $\Hcal^c$ as
$$
\Hcal^c := \left(\bigcap_{q: c_q = 0} H_0^q\right) \cap \left( \bigcap_{q: c_q = 1} H_1^q\right).
$$
Considering $Q=3$ tests and the configuration $c=(0, 1, 1)$, $\Hcal^c$ states that $H^1_0$ holds, but neither $H^2_0$ nor $H^3_0$.

Now, considering two complementary subsets $\Ccal_0$ and $\Ccal_1$ of $\Ccal$ (that is: $\Ccal_0 \cap \Ccal_1 = \varnothing$, $\Ccal_0 \cup \Ccal_1 = \Ccal$), we define the {\sl composed null} and {\sl alternative} hypotheses $\Hcal_0$ and $\Hcal_1$ as
$$
\Hcal_0 := \bigcup_{c \in \Ccal_0}\Hcal^c, \qquad
\Hcal_1 := \bigcup_{c \in \Ccal_1}\Hcal^c.
$$
As an example, in the case where $Q=3$ the Intersection Union test corresponds to the case where $\Ccal_1$ reduces to the configuration $c_{\max} = (1, 1, 1)$ and $\Ccal_0$ to the union of all others: $\Ccal_0 = \Ccal \setminus \{c_{\max}\}$. Alternatively, if one aims at detecting items such that hypothesis $\Hcal_0 = \{\text{less than two $H_1$ hypotheses hold among the three}\}$ is rejected, then one can define $\Ccal_1 = \{(1, 1, 0), (1, 0, 1), (0, 1, 1), (1, 1, 1)\}$.

In the sequel, we consider an experiment involving $n$ items (e.g. genes or markers), for which one wants to test $\Hcal_0$ versus $\Hcal_1$. We will denote by $\Hcal_{0i}$ the null composed hypothesis for item $i$ ($1 \leq i \leq n$) and similarly for $H^q_{0i}$, $\Hcal^c_i$ and so on. 

\paragraph{Joint mixture model.} 
Assume that $Q$ tests have been performed on each of the $n$ items and denote by $P_i^q$ the $p$-value obtained for test $q$ on item $i$. We note $P_i := (P^1_i, \dots, P^Q_i)$ the $p$-value profile of item $i$. Let us further define $Z^q_i$ the binary variable being 0 if $H^q_{0i}$ holds and 1 if $H^q_{1i}$ holds. A vector $Z_i := (Z^1_i, \dots Z^Q_i) \in \Ccal$ is hence associated with each item. Assuming that the items are independent, each $p$-value profile arises from a mixture model with $2^Q$ components defined as follows:
\begin{itemize}
\item the vectors $\{Z_i\}_{1 \leq i \leq n}$ are drawn independently within $\Ccal$ with probabilities $w_c = \Pr\{Z_i = c\}$, 
\item the $p$-value profiles $\{P_i\}_{1 \leq i \leq n}$ are drawn independently conditionally on the $Z_i$'s with distribution $\psi^c$:
$$
\left(P_i \mid Z_i = c\right) \sim \psi^c.
$$
\end{itemize}
So, the vectors $P_i$ are all independent and arise from the same multivariate mixture distribution
\begin{equation} \label{eq:multiMixtPi}
P_i \sim \sum_{c \in \Ccal} w_c \psi^c.
\end{equation}

\paragraph{A classification point of view.} 
In this framework, the $\Hcal_{0}$ items  correspond to items for which $Z_i \in \Ccal_0$, whereas the $\Hcal_{1}$ items are the ones for which $Z_i \in \Ccal_1$. Model \eqref{eq:multiMixtPi} can be rewritten as
$$
P_i \sim W_0 \psi_0 + W_1  \psi_1
$$
where $W_0 := \sum_{c \in \Ccal_0} w_c$, $\psi_0 := W_0^{-1}\sum_{c \in \Ccal_0} w_c \psi^c$ and respectively for $W_1$ and $\psi_1$. Hence, following \cite{ETS01}, we may turn the question of significance into a classification problem and focus on the evaluation of the conditional probability
\begin{equation} \label{eq:condProb}
\tau_i := \Pr\{Z_i \in \Ccal_1 \mid P_i\} = \frac{W_1 \psi_1(P_i)}{W_0 \psi_0(P_i) + W_1  \psi_1(P_i)},
\end{equation}
which is also known as the local FDR \cite{ABD04,ETS01,Str08,GRC09}.

\paragraph{About intersection-union.} 
In the particular case of the IUT, a major difference between the classical approach  and the one presented here is that the natural criterion to select the items for which $\Hcal_{1}$ is likely to hold are the posterior probabilities and not the maximal $p$-value $P^{\max}_i = \max_q P^q_i$.
This of course changes the ranking of the items in terms of significance (see Section \ref{sec:Ranking}). As will be shown in Section \ref{sec:Simul} and \ref{sec:Appli}, this modified ranking has a huge impact on the power of the overall procedure. As mentioned earlier the formalism we adopt enables one to consider more complex composed hypotheses than the IUT, and the same beneficial ranking strategy will be applied whatever the composed hypothesis tested. 

\paragraph{Modeling the $\psi^c$.} The mixture model \eqref{eq:multiMixtPi} involves $2^Q$ multivariate distributions $\psi^c$ that need to be estimated, which may become quite cumbersome when $Q$ becomes large. To achieve this task, we will assume that all distributions have the following product form:
\begin{equation} \label{eq:productForm}
\psi^c(P_i) = \prod_{q: c_q = 0} f^q_0(P^q_i) \prod_{q: c_q = 1} f^q_1(P^q_i), 
\end{equation}
so that only the $2Q$ univariate distributions $f^1_0, \dots f^Q_0, f^1_1, \dots f^Q_1$ need to be estimated. We emphasize that this product form does {\sl not} imply that the $p$-values $P^q_i$ are independent from one test to another, because no restriction is imposed on the proportions $w_c$, that control the joint distribution of $(Z^1_i, \dots Z^Q_i)$. \SR{}{Equation \eqref{eq:productForm} only means that the $Q$ $p$-values are independent conditionally on the configuration associated with entity $i$; they are not supposed to be marginally independent (See Appendix \ref{app:psiC}).}

\section{Inference \label{sec:Infer}} 

The procedure we propose can be summarized into 3 steps:
\begin{enumerate}
\item Fit a mixture model on each set of (\SR{}{probit-}transformed) $p$-values $\{P^q_i\}_{1 \leq i \leq n}$ to get an estimate of each \SR{(transformed)}{} alternative distribution $f^q_1$;
\item Estimate the proportion $w_c$ of each configuration $c$ using an EM algorithm and deduce the estimates of the conditional probabilities of interest $\tau_i$;
\item Rank the items according to the $\widehat{\tau}_i$ and compute an estimate of the false discovery rate to control for multiple testing.
\end{enumerate}

\subsection{Marginal distributions} 

The marginal distribution of the $p$-values $P^q_i$ associated with the $q$-th test can be deduced from Model \eqref{eq:multiMixtPi} combined with \eqref{eq:productForm}. One has 
\begin{equation} \label{eq:uniMixtPiq}
P^q_i \sim \pi^q_0 f^q_0 + (1- \pi^q_0) f^q_1,
\end{equation}
where $f^q_0$ is the distribution of $P^q_i$ conditional on $Z^q_i = 0$ and $f^q_1$ its distribution conditional on $Z^q_i = 1$. The proportion $\pi^q_0$ is a function of the proportions $w_c$. Now, because each $P^q_i$ is a $p$-value, its null distribution (i.e. its distribution conditional on $Z^q_i = 0$) is uniform over $(0, 1)$. Because this holds for each test, we have that $f^q_0(P^q_i) \equiv 1$ for all $i$'s and $q$'s.

\paragraph{Fitting marginal mixture models.} 
The mixture model \eqref{eq:uniMixtPiq} has received a lot of attention in the past because of its very specific form, one of its components being completely determined (and uniform). This specificity entails a series of nice properties. For example, \cite{Sto02} introduced a very simple yet consistent estimate of the null proportion $\pi^q_0$, namely
$$
\widehat{\pi}^q_0 = \SR{2 n^{-1}|\{i: P^q_i > .5\}|.}{[n (1 - \lambda)]^{-1}|\{i: P^q_i > \lambda\}|,}
$$
\SR{}{where we set $\lambda = .5$, which amounts to assume that the alternative distribution $f_1^q$ has no mass above $.5$.}
Given such an estimate, \cite{RBD07} showed that the alternative density can be estimated in a non-parametric way. 
\SR{Indeed taking $X^q_i$ as the negative probit transform of $P^q_i$: $X^q_i = -\Phi^{-1}(P^q_i)$ (where $\Phi$ stands for the standard Gaussian cdf and $\phi$ for its pdf)Model \eqref{eq:uniMixtPiq} becomes}{To this aim, they resort to the negative probit transform: $X^q_i = -\Phi^{-1}(P^q_i)$ (where $\Phi$ stands for the standard Gaussian cdf and $\phi$ for its pdf) to better focus on the distribution tails (see also \cite{Efron2008,MBB06,MDA05,RBD07}). Model \eqref{eq:uniMixtPiq} thus becomes} 
$$
X^q_i \sim \pi^q_0 \phi + (1- \pi^q_0) g^q_1\SR{.}{,}
$$ 
\SR{Because $\phi$ is known and $\pi^q_0$ has a prior estimate, a kernel estimate of $g^q_1$ can be defined as}{where the null pdf is known to be $\phi$ and where $\pi^q_0$ has a prior estimate, so a kernel estimate of $g^q_1$ can be defined as}
\begin{equation} \label{eq:g-hat}
\widehat{g}^q_1(x) = \sum_{i=1}^n \widehat{\tau}^q_i K_h(x - X_i) \left/ \sum_{i=1}^n \widehat{\tau}^q_i \right.
\end{equation}
where $K_h$ is a kernel function (with width $h$) and where
\begin{equation} \label{eq:tau}
\widehat{\tau}^q_i = \frac{(1- \widehat{\pi}^q_0) \widehat{g}^q_1(X^q_i)}{\widehat{\pi}^q_0 \phi(X^q_i) + (1- \pi^q_0) \widehat{g}^q_1(X^q_i)}.
\end{equation}
\cite{RBD07} showed that there exists a unique set of $\{\widehat{\tau}^q_i\}$ satisfying \SR{both equations that}{both \eqref{eq:g-hat} and \eqref{eq:tau}, which} can be found using a fix-point algorithm. 

In practice one needs to choose both the kernel function and its bandwidth $h$. In this article we used a Gaussian kernel function whose bandwidth can be tuned adaptively from the data  using cross-validation \cite{Chacon2018}. Both the Gaussian kernel density estimation and the cross-validation tuning are implemented in the $kde$ function of R package \texttt{ks} \cite{Duong2007}. 

\subsection{Configuration proportions} 

Likewise Model \eqref{eq:uniMixtPiq}, Model \eqref{eq:multiMixtPi} can be translated into a mixture model for the $X_i = (X^1_i, \dots X^Q_i)$, with same proportions $w_c$ but replacing each $f^q_0$ with $\phi$ and $f^q_1$ with $g^q_1$, 
namely
$$
X_i \sim \sum_{c \in \Ccal} w_c \gamma^c, \qquad 
\gamma^c(X_i) = \prod_{q: c_q=0} \phi(X^q_i) \prod_{q: c_q=1} g^q_1(X^q_i).
$$
The estimates $\widehat{g}_1^q$ introduced in the previous section directly provide us with estimates for the $\widehat{\gamma}^c$'s that can be plugged into the mixture, so that the only quantities to infer are the weights $w_c$. This inference can be efficiently performed using a standard EM thanks to closed-form expressions for the quantities to estimate at each step:
\begin{align*}
\text{E step:} & & \widehat{\Pr}\{Z_i = c \mid X_i\} 
& = \widehat{w}_c \widehat{\gamma}^c(X_i) \left/ \sum_{c' \in \Ccal} \widehat{w}_{c'} \widehat{\gamma}^{c'}(X_i) \right. , \\
\text{M step:} & &  \widehat{w}_c & = n^{-1} \sum_i \widehat{\Pr}\{Z_i = c \mid X_i\}.
\end{align*}

\subsection{Ranking and error control} \label{sec:Ranking}

As an important by product, the algorithm provides one with estimates of the conditional probabilities \eqref{eq:condProb} as
$$
\widehat{\tau}_i = \sum_{c \in \Ccal_1} \widehat{\Pr}\{Z_i = c \mid X_i\},
$$
according to which one can rank the items $1 \leq i \leq n$ so that
$$
\widehat{\tau}_1 > \widehat{\tau}_2 > ... > \widehat{\tau}_n.
$$
\SR{The conditional probabilities $\widehat{\tau}_i$ also provide a direct way to}{Following \cite{MDA05} (and \cite{MBB06,RBD07}), we use the conditional probabilities $\widehat{\tau}_i$ to} estimate the false discovery rate when thresholding at a given level $t$:
$$
\widehat{FDR}(t) = 1 - \frac1{N(t)} \sum_{i: \widehat{\tau}_i > t} \widehat{\tau}_i, 
\qquad N(t) = |\{i: \widehat{\tau}_i > t\}|.
$$
Consequently, threshold $t$ can be calibrated to control the \SR{T1}{type-I} error rate at a nominal level by setting 
$$
\hat{t} = \min_{\{ t: \widehat{FDR}(t) \leq \alpha\}} t.
$$ 
Note that the $\widehat{\tau}_i$'s are used twice, to rank the items and to estimate the FDR. Each of these two usages are investigated in the next section.\\

The whole strategy is called the QCH (Query of Composed Hypotheses) procedure hereafter. We emphasize that QCH already has two attractive features. First, because the inference steps 1 and 2 do not require the knowledge of the composed hypothesis to be tested, once the model is fitted one can query any number of composed hypotheses without any additional computational burden. Second, because the number of components in mixture model \citep{MaP00} is directly deduced from the number of $p$-value sets, the procedure comes with no parameter to tune for the user.    

\section{Simulations \label{sec:Simul}} 
In this section the performance of the QCH method is evaluated and compared to those of two \TMH{baseline}{} procedures \TMH{classically}{previously} considered for the IUT:
\begin{itemize}
\item The Pmax procedure consists in considering for each gene $i$ its associated maximum $p$-value $P_i^{\max} = \max(P_i^1,...,P_i^Q)$ as both the test statistic, for ranking, and the associated $p$-value, for false positive (FP) control. Once $P_i^{\max}$ is computed a multiple testing procedure is applied. Here we applied the Benjamini-Hochberg (BH, \cite{Benjamini1995}) procedure for FDR control. \TMH{}{This procedure corresponds to the IUT procedure described in \cite{Zhong2019}.}
\item The FDR set intersection procedure (called hereafter IntersectFDR) consists in applying a FDR control procedure \SR{}{(namely, the BH procedure)} to each of the $Q$ pvalue sets. The lists of items for which $H_0^q$ has been rejected are then intersected. \TMH{}{This corresponds to the "list crossing strategy" presented in the Introduction section (see e.g. \cite{Conway2017}).} 
\end{itemize}
\SR{}{As stated in the Model section the set $\Ccal_1$ of the corresponding QCH procedure reduces to the $c_{max}$ configuration that consists only of 1's.}
All the analyses presented in this section were performed with a nominal \SR{T1}{type-I} error rate level of $\alpha=5$\%.

We first consider a scenario where $Q$ sets of $n$ $p$-values are generated as follows. First, the proportion of $H_0$ hypotheses in each set is drawn according to a Dirichlet distribution $\mathcal{D}(8,2)$. 
\SR{The configuration proportions are deduced from these initial $H_0$ proportions, then the proportion corresponding to the full $H_1$ configuration is boosted to achieve at least 3\%. This boosting step ensures a minimal representation of the configuration one wants to detect. }{The configuration proportions are deduced from these initial $H_0$ proportions, as well as the proportion of the corresponding full $H_1$. The sampling was repeated to get a  full $H_1$ proportion of at least 3\%. This ensures a minimal representation of the configuration one wants to detect. }
Note that it also yields non independent $p$-values across the test, as the probability to be under any configuration is not the product of the corresponding marginal probabilities. The configuration memberships $Z_i$ are then independently drawn according to Model \eqref{eq:multiMixtPi}. The test statistics $T_{iq}\sim \mathcal{N}(\mu_{iq},1)$ are drawn independently, where $\mu_{iq}=0$ if $H_0$ holds for item $i$ in condition $q$, and $\mu_{iq}=2$ otherwise. Lastly these test-statistics are translated into $p$-values. Several values of $n$ ($10^4$, $10^5$, $10^6$) and $Q$ (2, 4, 8) are considered, and for a given parameter setting $(n,Q)$ \TMH{20}{100} $p$-value matrices were generated. This scenario is called "Equal Power" hereafter, since the deviation $\mu_{iq}$ under $H_1$ are the same for all $p$-value sets.

\begin{table*}\label{TableEqualPower2}
\centering
\begin{tabular}[t]{rrllllll}
\toprule
\multicolumn{1}{c}{ } & \multicolumn{1}{c}{ } & \multicolumn{2}{c}{Pmax\_BH} & \multicolumn{2}{c}{IntersectFDR} & \multicolumn{2}{c}{QCH} \\
\cmidrule(l{3pt}r{3pt}){3-4} \cmidrule(l{3pt}r{3pt}){5-6} \cmidrule(l{3pt}r{3pt}){7-8}
NbObs & Q & FDR & Power & FDR & Power & FDR & Power\\
\midrule
\rowcolor{gray!6}  1e+04 & 2 & 0 (0) & 0 (0.001) & 0.059 (0.089) & 0.034 (0.025) & 0.054 (0.1) & 0.031 (0.022)\\
1e+04 & 4 & 0 (0) & 0 (0) & 0 (0) & 0.001 (0.002) & 0.024 (0.038) & 0.095 (0.086)\\
\rowcolor{gray!6}  1e+04 & 8 & 0 (0) & 0 (0) & 0 (0) & 0 (0) & 0.004 (0.007) & 0.363 (0.134)\\
1e+05 & 2 & 0 (0) & 0 (0) & 0.059 (0.031) & 0.031 (0.023) & 0.044 (0.025) & 0.027 (0.018)\\
\rowcolor{gray!6}  1e+05 & 4 & 0 (0) & 0 (0) & 0.012 (0.049) & 0.001 (0.001) & 0.032 (0.022) & 0.09 (0.082)\\
1e+05 & 8 & 0 (0) & 0 (0) & 0 (0) & 0 (0) & 0.004 (0.002) & 0.364 (0.112)\\
\rowcolor{gray!6}  1e+06 & 2 & 0 (0) & 0 (0) & 0.058 (0.022) & 0.032 (0.023) & 0.047 (0.008) & 0.027 (0.02)\\
1e+06 & 4 & 0 (0) & 0 (0) & 0.009 (0.018) & 0.001 (0.001) & 0.03 (0.007) & 0.094 (0.08)\\
\rowcolor{gray!6}  1e+06 & 8 & 0 (0) & 0 (0) & 0 (0) & 0 (0) & 0.003 (0.001) & 0.349 (0.105)\\
\bottomrule
\end{tabular}
\caption{Performance of the 3 procedures for the "Equal Power" scenario. For each procedure the FDR and Power (averaged over 100 runs) are displayed for different settings. Numbers in brackets correspond to standard errors.}
\end{table*}

\begin{figure}
    \centering
    \begin{tabular}{cc}
        \includegraphics[height=5cm]{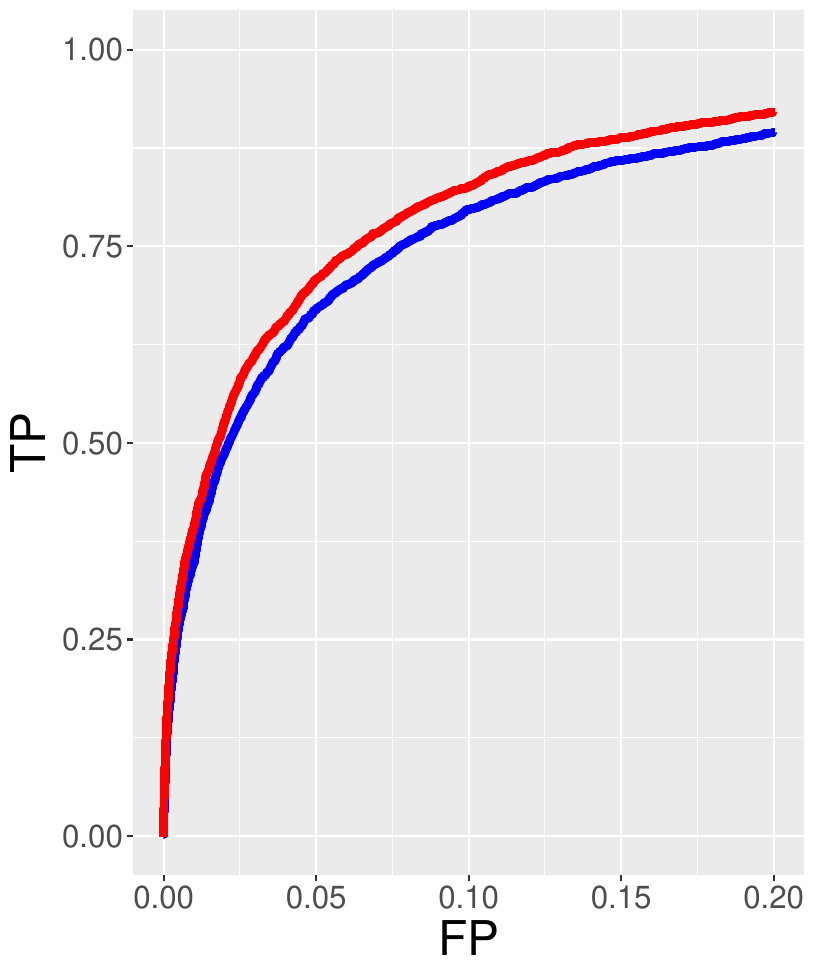} &  
        \includegraphics[height=5cm]{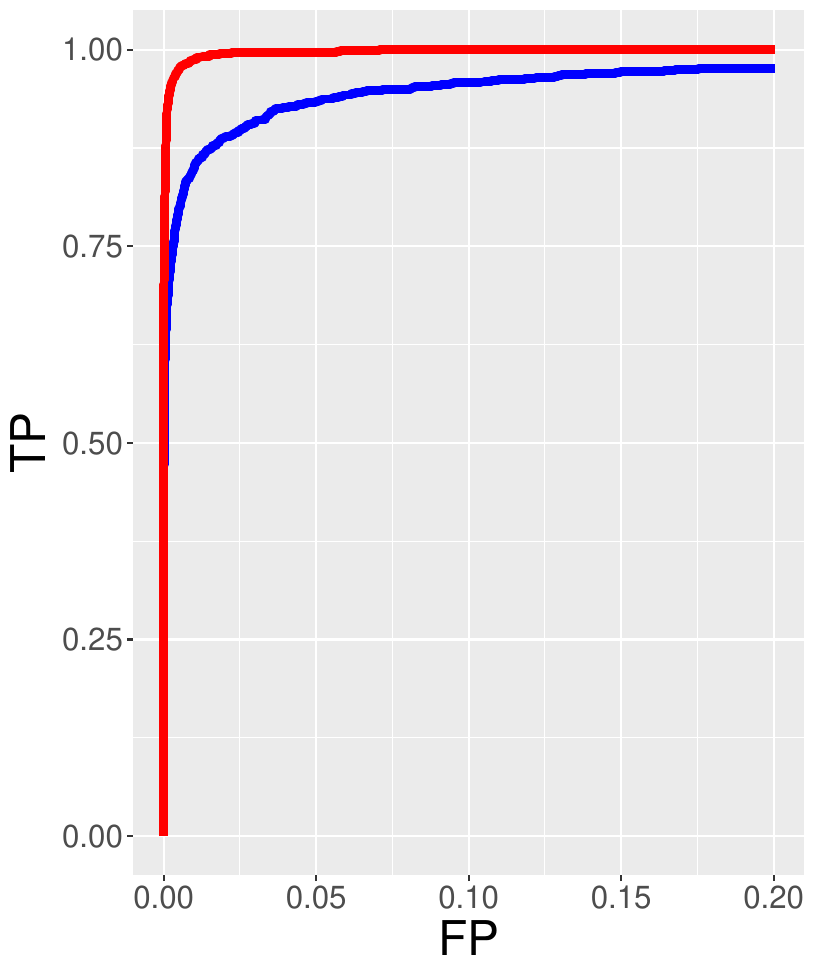}
    \end{tabular}
    \caption{\textbf{Left:} ROC curves for the Pmax (blue) and QCH (red) procedures, when $Q=2$. The x-axis corresponds displays the FP rate and the y-axis the TP rate. \textbf{Right:} same graph with $Q=8$.}
    \label{Figure_ROC}
\end{figure}

Figure \ref{Figure_ROC} displays the ROC curves of the Pmax and QCH procedures, computed on a single simulation with $n=10^5$ and $Q=2$  or 8. The x-axis (resp. the y-axis) corresponds to the FP rate (resp. TP = true positive rate) computed for all possible cutoffs on either $P_i^{\max}$ or the posterior probabilities $\widehat{\tau}_i$. When $Q=2$ the two procedures are roughly equivalent in terms of ranking, with a slight advantage for QCH. When $Q=8$ QCH significantly outperforms Pmax and reaches an AUC close to 1. This behavior is even more pronounced for larger values of $n$ (not shown), and show that \TMH{$P^{\max}$}{using $P^{\max}$ as a test statistic} is only relevant for small values of $Q$ (typically 2 or 3). \TMH{whereas the ranking on the QCH posterior probabilities always performs better than Pmax, and improves when $Q$ increases}{Alternatively the ranking on the QCH posterior probabilities always performs better than Pmax, and the higher $Q$ the higher the gap in terms of AUC.}. \TMH{}{Also note that one cannot easily compare the performance of the IntersectFDR procedure to the ones of Pmax and QCH since IntersectFDR is based on list crossing rather than selecting items based on an explicit test statistic.}

Table \ref{TableEqualPower2} provides some additional insight in terms of comparison between methods. First, one can observe that the \SR{T1}{type-I} error rate is always guaranteed with Pmax and QCH but not when intersecting FDR lists. Importantly, whatever $Q$ the power of Pmax is always close to 0 whereas the power of QCH is not. Since there are little differences between Pmax and QCH in terms of ranking when $Q=2$, it means that the multiple testing procedure for Pmax is conservative, i.e. the maximum $p$-value is a relevant test statistic, but should not be used directly as the $p$-value of the test. One can also observe that the power of the two baseline methods decreases whenever $n$ or $Q$ increases whereas QCH displays an opposite behavior. Indeed increasing the number of items $n$ yields more precision when fitting the mixture model and therefore sharper posterior probabilities. 

\begin{table*}
\centering
\centering
\begin{tabular}[t]{rrllllll}
\toprule
\multicolumn{1}{c}{ } & \multicolumn{1}{c}{ } & \multicolumn{2}{c}{Pmax\_BH} & \multicolumn{2}{c}{IntersectFDR} & \multicolumn{2}{c}{QCH} \\
\cmidrule(l{3pt}r{3pt}){3-4} \cmidrule(l{3pt}r{3pt}){5-6} \cmidrule(l{3pt}r{3pt}){7-8}
NbObs & Q & FDR & Power & FDR & Power & FDR & Power\\
\midrule
\rowcolor{gray!6}  1e+04 & 2 & 0 (0) & 0.001 (0.003) & 0.059 (0.039) & 0.134 (0.063) & 0.05 (0.044) & 0.138 (0.069)\\
1e+04 & 4 & 0 (0) & 0 (0) & 0.01 (0.02) & 0.127 (0.06) & 0.04 (0.016) & 0.713 (0.179)\\
\rowcolor{gray!6}  1e+04 & 8 & 0 (0) & 0 (0.001) & 0 (0) & 0.133 (0.065) & 0.028 (0.015) & 0.995 (0.021)\\
1e+05 & 2 & 0 (0) & 0 (0) & 0.056 (0.025) & 0.127 (0.056) & 0.048 (0.009) & 0.135 (0.052)\\
\rowcolor{gray!6}  1e+05 & 4 & 0 (0) & 0 (0) & 0.009 (0.01) & 0.124 (0.049) & 0.037 (0.008) & 0.696 (0.163)\\
1e+05 & 8 & 0 (0) & 0 (0) & 0 (0.001) & 0.136 (0.054) & 0.024 (0.007) & 1 (0.001)\\
\rowcolor{gray!6}  1e+06 & 2 & 0.001 (0.013) & 0 (0.002) & 0.062 (0.019) & 0.137 (0.059) & 0.05 (0.003) & 0.137 (0.058)\\
1e+06 & 4 & 0 (0) & 0 (0) & 0.009 (0.011) & 0.126 (0.058) & 0.038 (0.007) & 0.705 (0.172)\\
\rowcolor{gray!6}  1e+06 & 8 & 0 (0) & 0 (0) & 0 (0) & 0.129 (0.053) & 0.023 (0.003) & 1 (0)\\
\bottomrule
\end{tabular}
\caption{\label{TableLinearPower2} Performance of the 3 procedures for the "Linear Power" scenario. For each procedure the FDR and Power (averaged over 100 runs) are displayed for different settings. Numbers in brackets correspond to standard errors.} \label{Table_InequalPower}
\end{table*}

In a second scenario called \TMH{}{"Linear power"} $p$-values sets were generated the same way as in the previous case, except for the fact that the deviation parameter $\mu_{iq}$ is 0 if $H_0$ holds for item $i$ in condition $q$, and $\mu_{iq}=\mu_{q}$ otherwise, where \TMH{}{$\mu_q=q+1$} for $q=1,...,Q$, i.e. the statistical power associated to set $q$ increases with $q$. The performances of the 3 procedures are displayed in Table \ref{Table_InequalPower}. One can observe that neither the Pmax nor the IntersectFDR procedure get any benefit from the fact that the distinction between $H_0$ and $H_1$ is easy for some $p$-value sets. To the contrary, QCH fully exploits this information, achieving power higher than 60\% when Pmax is at 0 and IntersectFDR is lower than 0.05. 

Lastly, we considered the composed alternative hypothesis 
$$
H_1:\{\text{Item } i \text{ is } H_1 \text{ in at least } Q-1 \text{ conditions}\}.
$$
The two procedures Pmax and FDRintersect can be empirically modified to handle such composed hypothesis as follows :
\begin{itemize}
\item rather than computing Pmax as the maximum $p$-value one can get the second largest $p$-value,
\item rather than intersecting all $Q$ FDR-lists one can consider all the combinations of $Q-1$ intersected lists among $Q$.
\end{itemize}
Note that these adaptations are feasible for some composed hypotheses but usually lead to a complete redefinition of the testing procedure that can become quite cumbersome, whereas no additional fitting is required for QCH. Indeed, for each new composed hypothesis to be tested only the posterior probabilities corresponding to the new hypothesis have to be computed. This comes without any additional computational cost since one only needs to sum up configuration posterior probabilities corresponding to the new hypothesis without does not require one to refit model \eqref{eq:multiMixtPi}. The results are provided in Table \ref{TableQm1LinearPower2} for the "Linear Power" scenario. In terms of performance the detection problem becomes easier for all procedures (since the proportion of $\mathcal{H}_1$ items is now much bigger). \TMH{}{Still, QCH consistently achieves the best performance and significantly outperforms its competitors as soon as $Q\geq4$, being close to a 100\% detection rate when $Q=8$, while being efficient at controlling the FDR rate at its nominal level. Additional results on alternative simulation settings can be found in Appendix.}

\begin{table*}
\centering
\centering
\begin{tabular}[t]{rrllllll}
\toprule
\multicolumn{1}{c}{ } & \multicolumn{1}{c}{ } & \multicolumn{2}{c}{Pmax\_BH} & \multicolumn{2}{c}{IntersectFDR} & \multicolumn{2}{c}{QCH} \\
\cmidrule(l{3pt}r{3pt}){3-4} \cmidrule(l{3pt}r{3pt}){5-6} \cmidrule(l{3pt}r{3pt}){7-8}
NbObs & Q & FDR & Power & FDR & Power & FDR & Power\\
\midrule
\rowcolor{gray!6}  1e+04 & 2 & 0.06 (0.012) & 0.624 (0.111) & 0.031 (0.006) & 0.519 (0.123) & 0.05 (0.004) & 0.619 (0.126)\\
1e+04 & 4 & 0.005 (0.004) & 0.262 (0.029) & 0.022 (0.015) & 0.411 (0.048) & 0.049 (0.007) & 0.733 (0.093)\\
\rowcolor{gray!6}  1e+04 & 8 & 0 (0) & 0.201 (0.018) & 0 (0.001) & 0.353 (0.037) & 0.036 (0.007) & 1 (0)\\
1e+05 & 2 & 0.063 (0.013) & 0.608 (0.116) & 0.032 (0.007) & 0.502 (0.127) & 0.05 (0.002) & 0.598 (0.133)\\
\rowcolor{gray!6}  1e+05 & 4 & 0.005 (0.004) & 0.265 (0.011) & 0.021 (0.013) & 0.412 (0.04) & 0.047 (0.003) & 0.727 (0.092)\\
1e+05 & 8 & 0 (0) & 0.199 (0.005) & 0 (0) & 0.365 (0.038) & 0.029 (0.004) & 1 (0.001)\\
\rowcolor{gray!6}  1e+06 & 2 & 0.062 (0.012) & 0.613 (0.11) & 0.032 (0.006) & 0.504 (0.122) & 0.05 (0.001) & 0.601 (0.128)\\
1e+06 & 4 & 0.005 (0.003) & 0.264 (0.009) & 0.022 (0.013) & 0.413 (0.035) & 0.047 (0.002) & 0.724 (0.088)\\
\rowcolor{gray!6}  1e+06 & 8 & 0 (0) & 0.199 (0.002) & 0 (0) & 0.356 (0.032) & 0.029 (0.004) & 1 (0.001)\\
\bottomrule
\end{tabular}
\caption{\label{TableQm1LinearPower2} Performance of the 3 procedures for the "Linear Power" scenario and the "at least $Q-1$ $H_1$ composed hypothesis. For each procedure the FDR and Power (averaged over 100 runs) are displayed for different settings. Numbers in brackets correspond to standard errors.}
\end{table*}

\section{Illustrations \label{sec:Appli}}  
\subsection{TSC dataset}

In our first application we consider the Tuberous Sclerosis Complex (TSC) dataset obtained from the Database of Genotypes and Phenotypes  website (dbGap: accession code phs001357.v1.p1). The dataset consists in 34 TSC and 7 control tissue samples, for which gene expression and/or methylation were quantified. A differential analysis was performed separately on the two types of omics data, leading to the characterization of 7,222 genes for expression and 273,215 Cpg sites for methylation. In order to combine the corresponding $p$-values sets, we considered pairs between a gene and a Cpg site, the pairs being constituted according to one of the following two strategies:
\begin{itemize}
\item pair a gene with the methylation sites  directly located at the gene position; this strategy resulted in 128,879 pairs (some Cpg sites being not directly linked to any gene) and is called "Direct" (for Direct vicinity) hereafter, 

\item pair a gene with any methylation site that is physically close to that gene due to chromatin interactions; this strategy resulted in 6,532,368 pairs and is called "HiC" hereafter,  

\end{itemize}

Depending on the strategy a same gene could be paired with several methylation sites and vice versa. Strategy (b) requires additional information about chromatin information that was obtained from HiC data obtained from Gene Expression Omnibus (GEO accession code GSM455133). 

The purpose of the data integration analysis is then to identify pairs that exhibit a significant differential effect on both expression and methylation. In terms of composed hypothesis this boils down to perform IUT: the configurations corresponding to $\mathcal{H}_0$ is $\Ccal_0=\{00, 01, 10\}$ and $\mathcal{H}_1$ is $\mathcal{C}_1=\{11\}$.
In addition to using QCH, we evaluated the Pmax procedure associated with a Benjamini-Hochberg correction. 

The results are summarized in Table \ref{tab:TSC}. Using the Direct strategy combined with the QCH procedure, 4,030 genes were classified as $\mathcal{H}_1$ (out of 1.3M significant combinations). As for the simulations studies the QCH procedure detected an increased number of pairs and genes compared with Pmax. Applying the HiC strategy resulted in significantly higher number of tested pairs but a lower number of identified genes. Interestingly, the list of $\mathcal{H}_1$ genes detected with the HiC strategy contains many candidates that were not detected using the Direct strategy. Among these candidates found with the HiC strategy only are TSC1 and TSC2 for which mutations are well known to be associated with the TSC disease. The QCH procedure also identified three additional genes (again with the HiC strategy only) that were not detected using Pmax, and whose combination with methylation sites in TSC1 gene was significant. These genes are LRP1B, PEX14 and CHD7. LRP1B is associated with renal angiomyolipoma (AML) and AML is observed in $75\%$ patients with TSC \cite{Wang2020}. PEX14 is known to interact with PEX5 for importing proteins to the peroxisomes \cite{Neuhaus2014}. Mutations in CHD7 have been found to increase risk of idiopathic autism \cite{Oroak2012}, which could suggest a link between monogenic disorder like TSC and idiopathic autism due to complex inheritance \cite{Gamsiz2015}: it has indeed been reported that TSC patients may develop abnormal neurons and cellular enlargement, making  difficult the differentiation between TSC and autistic symptoms in the setting of severe intellectual disability \cite{Takei2014}. 
In conclusion the QCH procedure combined with the HiC information detected genes that have been previously reported to have functional association with the disease in different studies. These genes may reveal more insight about the genetic architecture of TSC. 

\begin{table}
\begin{center}
\begin{tabular}{l|cccc}
\hline
Strategy & Total Pmax & Total QCH  & Extra Pmax  & Extra QCH \\
\hline
Direct & 3624 & 4030 & 4 & 410  \\
HiC  & 3501 & 3875 & 0 & 374 \\
\hline
\end{tabular}
\end{center}
\caption{Number of $\mathcal{H}_1$ genes identified through different pairing strategies and procedures (Pmax or QCH).\label{tab:TSC}} 
\end{table}

\subsection{Einkorn dataset}

We consider the Einkorn study of \cite{Bonnot2020}, in which the grain transcriptome was investigated in four nutritional treatments at different time points during grain development (see the original reference for a complete description of the experimental design). For each combination of a treatment and a timepoint 3 einkorn grains were harvested and gene expression was quantified through RNAseq. The results of the differential analyses are publicly available at \url{forgemia.inra.fr/GNet/einkorn_grain_transcriptome/-/tree/master/}.
Here we focus on the comparaison between the NmSm (control) treatment and the NpSm (enriched in Nitrate) treatment, compared at $Q=4$ different timepoints (T300, T400, T500, T600). The differential analysis involved $n=12,327$ transcripts and was performed at each timepoint. We extracted the results what were summarized into a $n\times Q$ $p$-value matrix. 

\begin{figure}
    \centering
    \includegraphics[scale=0.5]{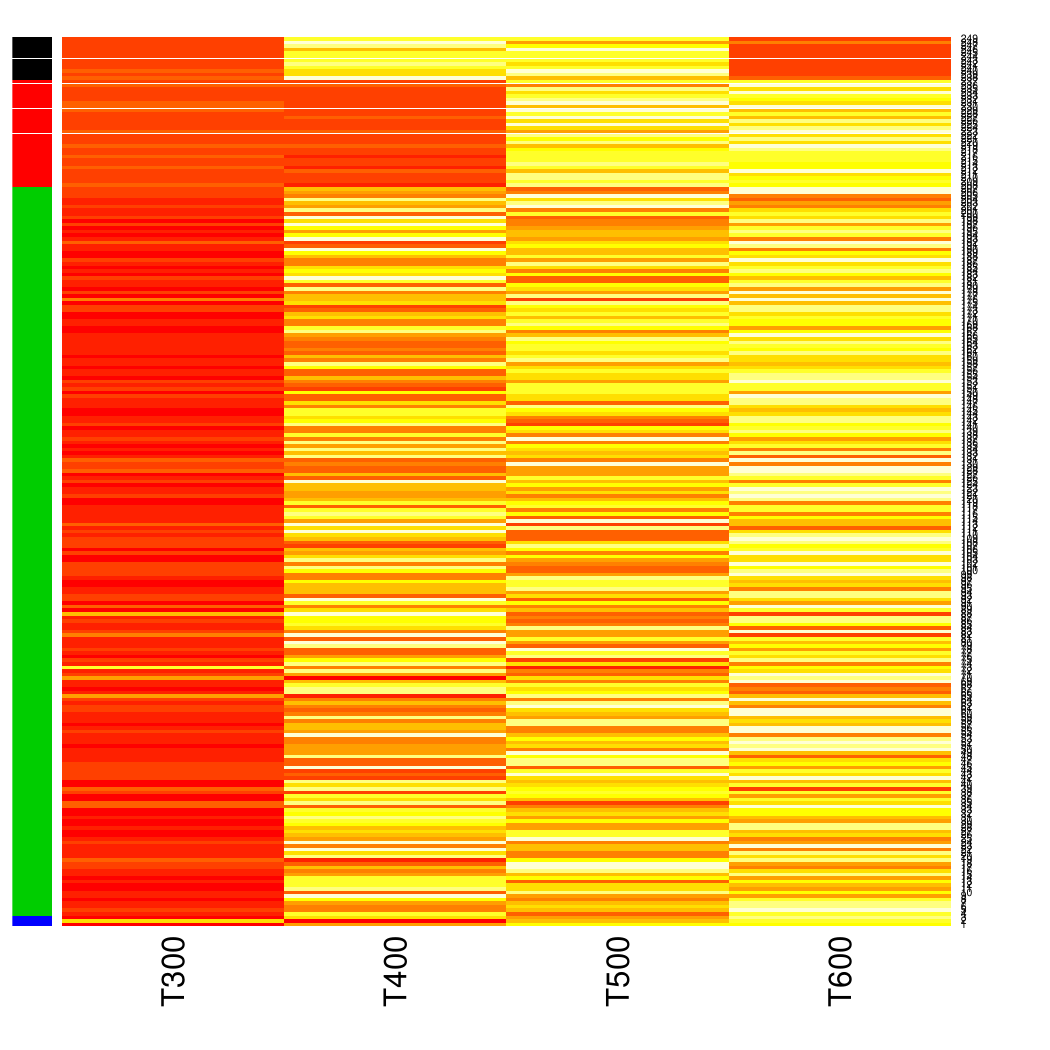}    
    \caption{Pvalue profiles of the $\mathcal{H}_1$ transcripts over time, after -log10 transformation. Colors on the left side correspond to the configurations (black=0110, red=0011, green=0111 and blue=1111) }
    \label{Figure_TimePointEinkorn}
\end{figure}

In this context, we consider the $\mathcal{H}_0$ composed null hypothesis 
$$
\mathcal{H}_{0i}:  \left\{
\begin{array}{c}
\text{transcript $i$ is not differentially expressed} \\
\text{at 2 consecutive timepoints}
\end{array} \right\}
$$
that corresponds to the following composed alternative subset:
$$
\mathcal{C}_1=\{1100, 0110, 0011, 1101, 1110, 1011, 0111, 1111\}.
$$
The results are displayed in Figure \ref{Figure_TimePointEinkorn} and Table \ref{TableEinkornTranscriptClass}. A total of 249 transcripts were declared significant for the composed alternative hypothesis at FDR nominal level of 5\%. These transcripts were further classified into a H-configuration according to their maximal posterior, resulting in a 4-class classification presented in Table \ref{TableEinkornTranscriptClass}. The table displays for each H-configuration the median value of the -log10($p$value) at the different timepoints over the transcripts belonging to the class. One can first observe that several H-configurations such as 1100 or 1110 are missing, i.e. no transcripts are DE in early times but not later. Alternatively, H-configurations 0110 and 0011 are well represented by 12 and 30 transcripts, respectively. These classes can be clearly observed on Figure \ref{Figure_TimePointEinkorn} that displayed the (-log10 transformed) $p$-value profiles of the 249 transcripts over time, classified according to their H-configuration (color level on the left side). One can observe the time-shifted effect of the transcripts belonging to these two classes. Identifying these transcripts provides valuable insight about the kinetics of the response to nitrate enrichment that could be incorporated in / confirmed by the network reconstruction to follow.

\begin{table}
\begin{center}
\begin{tabular}{lrrrrr}
\toprule
Config & NbTrspt & T300 & T400 & T500 & T600\\
\midrule
\rowcolor{gray!6}  0110 & 12 & 0.2225031 & 3.8242946 & 3.296769 & 0.3776435\\
\red{0011} & 30 & 0.2200245 & 0.1256215 & 4.479422 & 4.4642771\\
\rowcolor{gray!6}  \green{0111} & 204 & 0.2982042 & 2.4155116 & 2.851227 & 3.1877051\\
\blue{1111} & 3 & 11.9843371 & 14.9638330 & 15.251248 & 17.2187644\\
\bottomrule
\end{tabular}
\end{center}
\caption{\label{TableEinkornTranscriptClass}Description of the 4 H-config classes identified from the NmSm-NpSm comparison. For each class (in row) and each timepoint (in column) the median value of the -log10($p$value) is reported along with the number of detected transcripts (column NbTrspt).}
\end{table}

\section{Discussion \label{sec:Discus}}  
\paragraph{Summary.}
We introduced a versatile approach to assess composed hypotheses, that is any set of configurations of null and alternative hypotheses among $Q$ being tested. The approach relies on a multivariate mixture model that is fitted in an efficient manner so that very large omic datasets can be handled. The classification point-of-view we adopted yields a natural ranking of the items based on their associated posterior probabilities. These probabilities can also be used as local FDR estimates to compute (and control) the FDR. Note that in the present work we considered FDR control, but one could consider to directly control the local FDR \cite{Efron2008}, or alternatively to control more refined error rates such as the \SR{MFDR}{multiple FDR} developed in the context of multi-class mixture models \cite{Mary2013}. \\
\TMH{}{The Simulation section illustrated the gap between Pmax and QCH in terms of power. The poor performance of the Pmax procedure is due to the use of Pmax as both a test statistic (i.e. for ranking the items) and a p-value (i.e. the p-value associated to Pmax is assumed to be Pmax itself). Although this corresponds to what has been applied in practice \cite{Zhong2019}, one can observe that Pmax cannot be considered as a p-value since it is not uniformly distributed under the null (composed) hypothesis $\mathcal{H}_0$. Consequently a direct application of multiple testing correction procedures to Pmax will automatically lead to a conservative testing procedure. Although it may be feasible to improve on the current practice, note that i) finding the null distribution of Pmax and ii) extending the Pmax procedure to the more general question of testing a general composed hypothesis are two difficult tasks. The QCH methodology present here solves these two problems in an efficient way, in terms of power, FDR control and computational efficiency.}

\paragraph{Future works.}
The proposed methodology also provides information about the joint distributions of the latent variables $Z^q_i$, that is the probability that, say, both $H^1_i$ and $H^3_i$ hold, but not $H^2_i$. This distribution encodes the dependency structure between the tests. This available information should obviously be further carefully investigated as it provides insights about the way items (e.g. genes) respond to each combination of tested treatments.

Although the proposed model allows for dependency between the $p$-values through the distribution of the latent variables $Z_i$, conditional independence (i.e. independence within each configuration) is assumed to alleviate the computational burden. This assumption could be removed in various ways. A parametric form, such as a (probit-transformed) multivariate Gaussian with constrained variance matrix, could be considered for each joint distribution $\psi^c$. Alternatively, a non-parametric form could be preserved, using copulas to encode the conditional dependency structure.

\paragraph{Acknowledgements.}
SD and IM acknowledge the submitters of the TSC data to the dbGaP repository. \url{https://www.ncbi.nlm.nih.gov/projects/gap/cgi-bin/study.cgi?study_id=phs001357.v1.p1}

\paragraph{Funding}
This work has been supported by the Indo-French Center for Applied Mathematics (IFCAM) and the ”Investissement d’Avenir” project (Amaizing, ANR-10-BTBR-0001), Department of Biotechnology, Govt. of India, for their partial support to this study through SyMeC.\\

\paragraph{Availability and Implementation}
R codes to reproduce the Einkorn example are available on the personal webpage of the first author:
\url{https://www6.inrae.fr/mia-paris/Equipes/Membres/Tristan-Mary-Huard}. The QCH methodology is available in the \texttt{qch} package hosted on CRAN.



\newpage
\appendix
\section{Appendix \label{sec:Appendix}}  
\subsection{\SR{}{Conditional independence does not mean marginal independence} \label{app:psiC}}

We illustrate here the assumption underlying the product form of the distribution $\psi^c$ given in \eqref{eq:productForm}. More specifically, we remind that, although this product form amounts to assume that the $p$-values are conditionally independent, given configuration $Z_i = (Z_i^q)_{q = 1, \dots Q}$, they are not marginally independent because we do not assume that the status wrt to each hypothesis are independent.

To this aim, we consider $Q=2$ tests, so 4 configurations $c$ exists: $(0, 0)$, $(1, 0)$, $(0, 1)$ and $(1, 1)$. We set
$$
w_{(0, 0)} = .8, \quad w_{(1, 0)} = .05, \quad w_{(0, 1)} = .05, \quad w_{(1, 1)} = .1
$$
so that probability to be under $H_1$ is $.15$ for each test, but the probability to be under $H_1$ is $.15$ for both test is $.1 \gg .15^2$. The correlation between $Z_i^1$ and $Z_i^2$ is about $.6$, which corresponds to a situation were it is more likely to be $H_1$ for the second test when the entity is $H_1$ for the first test.

We simulated $n = 10^4$ entities, we draw $n$ corresponding configurations $Z_i$ according to the $w_c$ given above. For each entity $i$, for each test $q = 1, 2$, we then sampled independently the $p$-values $P_i^q$ from a uniform distribution if $Z_i^q = 1$ and to a Beta distribution $\text{B}(1, 20)$ if $Z_i^q = 1$.

\begin{figure*}[ht]
    \begin{center}
    \begin{tabular}{cc}
        \includegraphics[width=.33\textwidth]{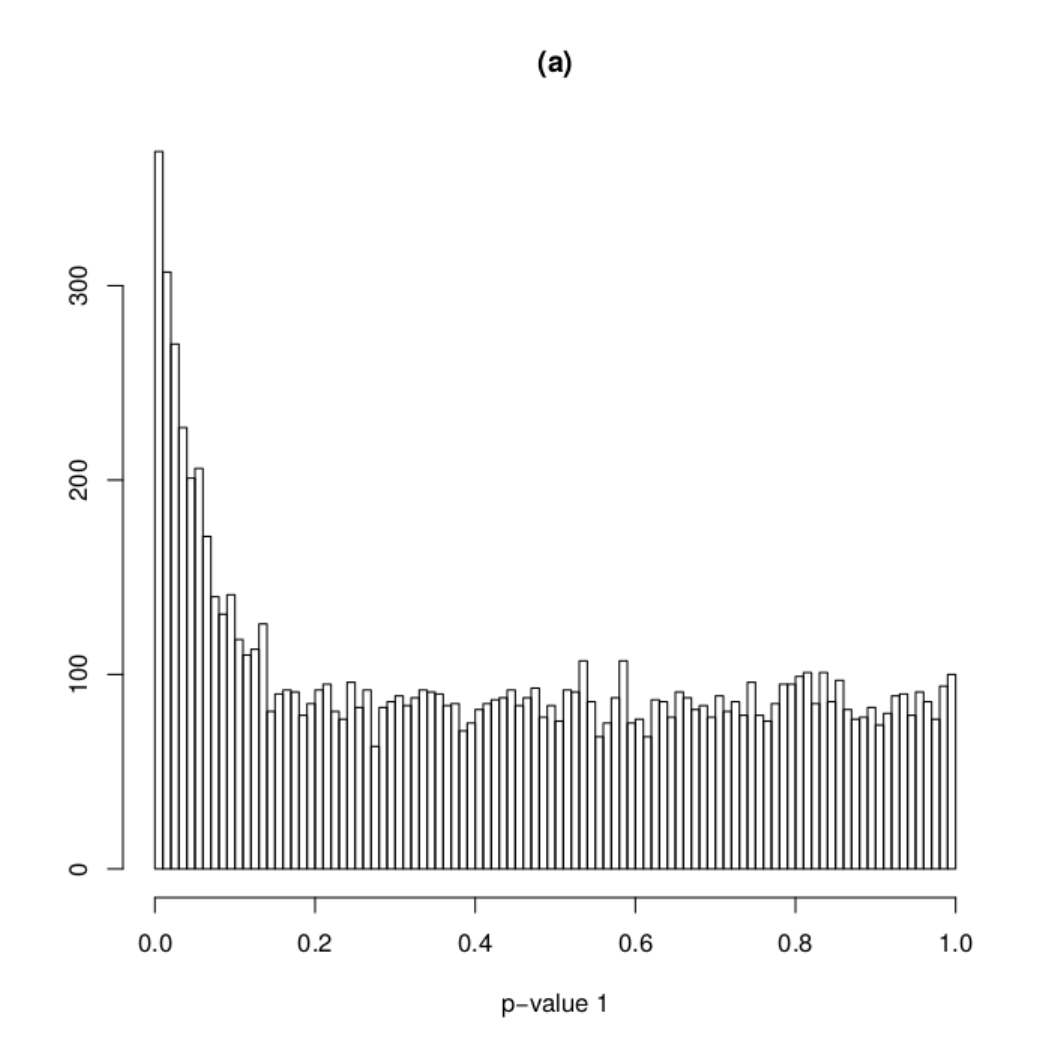} &
        \includegraphics[width=.33\textwidth]{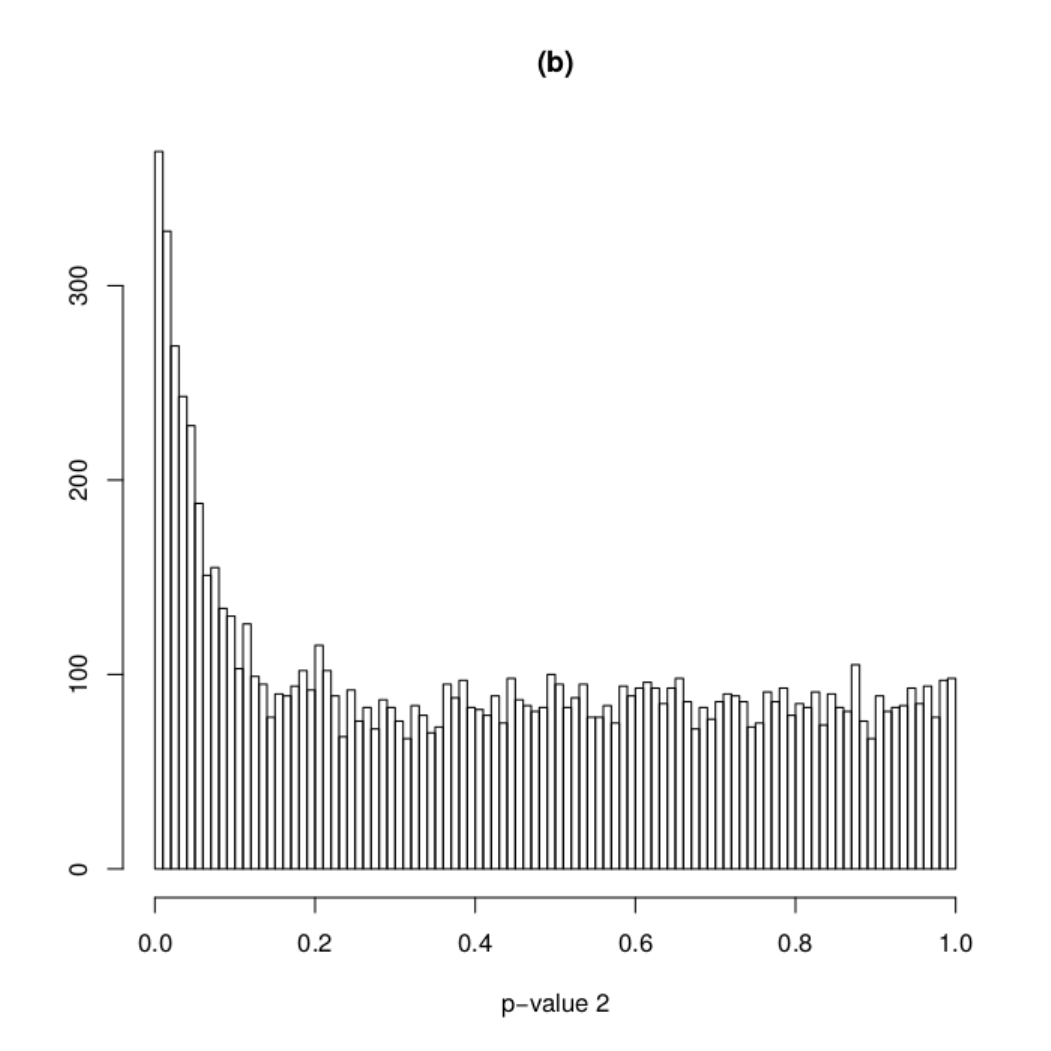} \\
        \includegraphics[width=.33\textwidth]{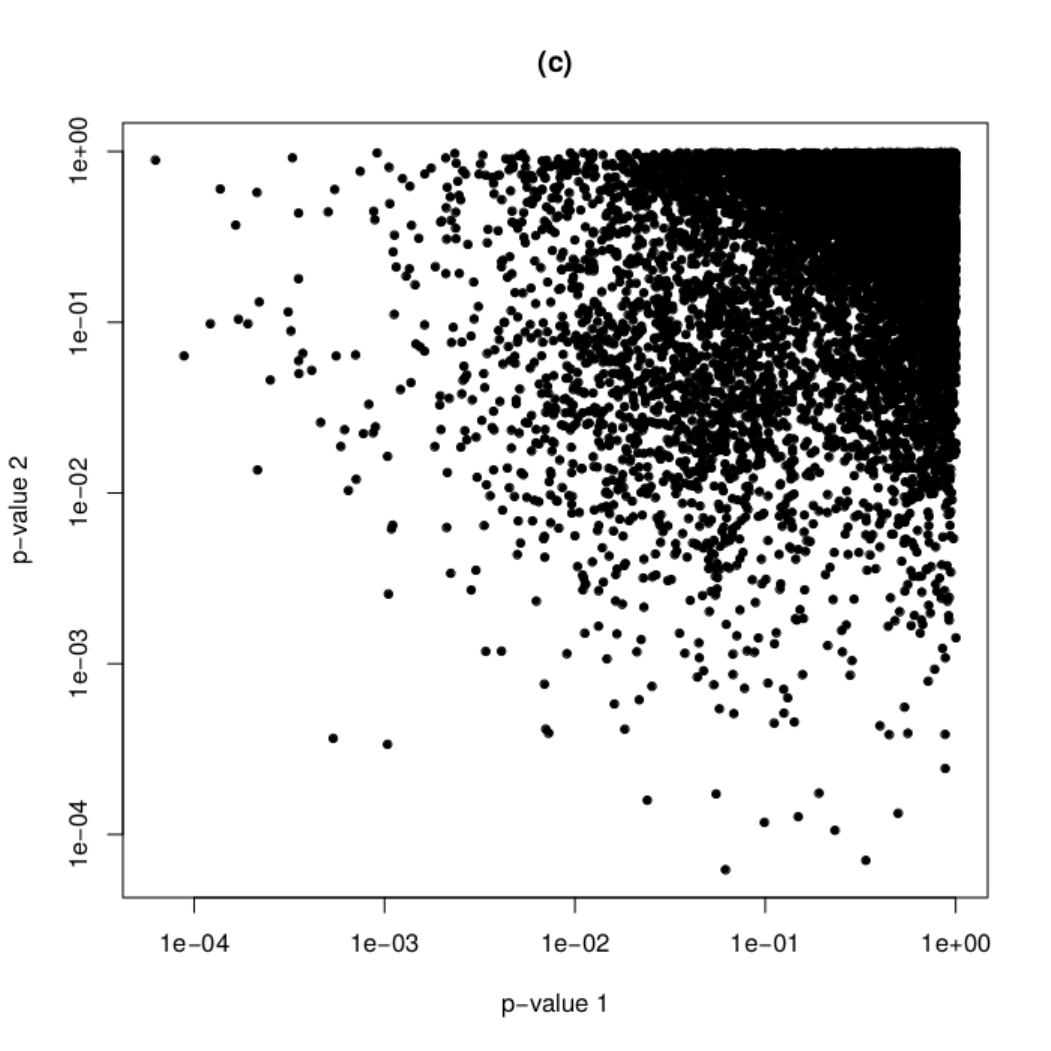} &
        \includegraphics[width=.33\textwidth]{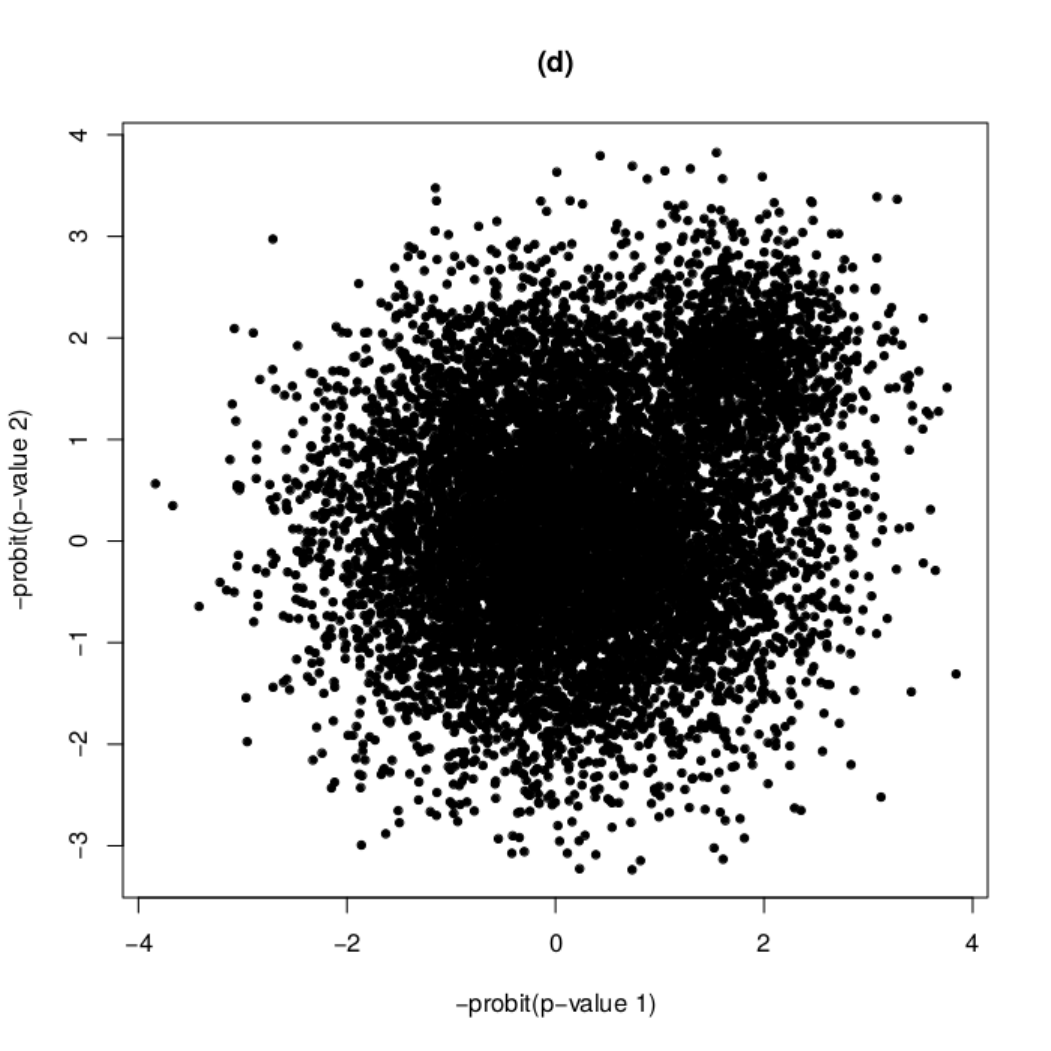} \\
    \end{tabular}
    \end{center}
    \caption{Simulated conditionally independent $p$-values. (a) Distribution of the $P_i^1$. (b) Distribution of the $P_i^2$. (c) Joint distribution of the $(P_i^1, P_i^2)$ (log-scale). (d) Joint distribution of the $(-\Phi^{-1}(P_i^1), -\Phi^{-1}(P_i^2))$ \label{fig:jointPvalues}}
\end{figure*}

Figure \ref{fig:jointPvalues} displays the results. The top panels (a) and (b) display typical distributions for the $p$-values of well calibrated test. The bottom panels display the joint distribution of these $p$-values (c) and of the negative-probit transforms (d). The observed correlation between the $P_i^1$ and the $P_i^2$ is $.16$ (and $.19$ for $(-\Phi^{-1}(P_i^1), -\Phi^{-1}(P_i^2))$). This correlation is entirely inherited from the this that exists between $Z_i^1$ and $Z_i^2$. This shows that the product form we adopt in \eqref{eq:productForm} does not prevent us to account for a biologically association that may exist between the two processes targeted  by the two tests respectively.

\subsection{Performance comparison}


\begin{table*}
\centering
\centering
\begin{tabular}[t]{rrllllll}
\toprule
\multicolumn{1}{c}{ } & \multicolumn{1}{c}{ } & \multicolumn{2}{c}{Pmax\_BH} & \multicolumn{2}{c}{IntersectFDR} & \multicolumn{2}{c}{QCH} \\
\cmidrule(l{3pt}r{3pt}){3-4} \cmidrule(l{3pt}r{3pt}){5-6} \cmidrule(l{3pt}r{3pt}){7-8}
NbObs & Q & FDR & Power & FDR & Power & FDR & Power\\
\midrule
\rowcolor{gray!6}  1e+04 & 2 & 0.004 (0.011) & 0.088 (0.068) & 0.057 (0.024) & 0.478 (0.081) & 0.046 (0.015) & 0.45 (0.073)\\
1e+04 & 4 & 0 (0) & 0 (0) & 0.01 (0.018) & 0.248 (0.058) & 0.048 (0.015) & 0.718 (0.131)\\
\rowcolor{gray!6}  1e+04 & 8 & 0 (0) & 0 (0) & 0 (0) & 0.06 (0.025) & 0.042 (0.013) & 0.992 (0.011)\\
1e+05 & 2 & 0.005 (0.006) & 0.084 (0.072) & 0.058 (0.018) & 0.475 (0.076) & 0.049 (0.006) & 0.454 (0.06)\\
\rowcolor{gray!6}  1e+05 & 4 & 0 (0) & 0 (0) & 0.009 (0.014) & 0.239 (0.066) & 0.045 (0.005) & 0.737 (0.149)\\
1e+05 & 8 & 0 (0) & 0 (0) & 0 (0) & 0.064 (0.023) & 0.042 (0.006) & 0.994 (0.008)\\
\rowcolor{gray!6}  1e+06 & 2 & 0.005 (0.003) & 0.082 (0.072) & 0.062 (0.017) & 0.48 (0.073) & 0.049 (0.002) & 0.447 (0.059)\\
1e+06 & 4 & 0 (0) & 0 (0) & 0.009 (0.011) & 0.241 (0.058) & 0.046 (0.003) & 0.732 (0.135)\\
\rowcolor{gray!6}  1e+06 & 8 & 0 (0) & 0 (0) & 0 (0) & 0.062 (0.022) & 0.047 (0.003) & 0.995 (0.009)\\
\bottomrule
\end{tabular}
\caption{All tests significant, Delta Equal, Effect size 3}
\end{table*}


\begin{table*}
\centering
\centering
\begin{tabular}[t]{rrllllll}
\toprule
\multicolumn{1}{c}{ } & \multicolumn{1}{c}{ } & \multicolumn{2}{c}{Pmax\_BH} & \multicolumn{2}{c}{IntersectFDR} & \multicolumn{2}{c}{QCH} \\
\cmidrule(l{3pt}r{3pt}){3-4} \cmidrule(l{3pt}r{3pt}){5-6} \cmidrule(l{3pt}r{3pt}){7-8}
NbObs & Q & FDR & Power & FDR & Power & FDR & Power\\
\midrule
\rowcolor{gray!6}  1e+04 & 2 & 0.01 (0.01) & 0.333 (0.098) & 0.062 (0.023) & 0.651 (0.081) & 0.049 (0.014) & 0.653 (0.074)\\
1e+04 & 4 & 0.001 (0.004) & 0.24 (0.043) & 0.011 (0.018) & 0.686 (0.078) & 0.067 (0.023) & 0.966 (0.043)\\
\rowcolor{gray!6}  1e+04 & 8 & 0 (0) & 0.236 (0.048) & 0 (0) & 0.663 (0.091) & 0.049 (0.006) & 1 (0)\\
1e+05 & 2 & 0.01 (0.006) & 0.332 (0.097) & 0.059 (0.021) & 0.65 (0.083) & 0.05 (0.004) & 0.67 (0.077)\\
\rowcolor{gray!6}  1e+05 & 4 & 0 (0.001) & 0.241 (0.017) & 0.008 (0.011) & 0.656 (0.078) & 0.057 (0.009) & 0.97 (0.043)\\
1e+05 & 8 & 0 (0) & 0.241 (0.016) & 0 (0) & 0.656 (0.071) & 0.048 (0.004) & 1 (0)\\
\rowcolor{gray!6}  1e+06 & 2 & 0.01 (0.005) & 0.336 (0.099) & 0.057 (0.019) & 0.645 (0.083) & 0.05 (0.001) & 0.665 (0.072)\\
1e+06 & 4 & 0.001 (0.001) & 0.239 (0.005) & 0.012 (0.016) & 0.667 (0.075) & 0.056 (0.006) & 0.958 (0.046)\\
\rowcolor{gray!6}  1e+06 & 8 & 0 (0) & 0.239 (0.005) & 0 (0) & 0.663 (0.086) & 0.049 (0.002) & 1 (0)\\
\bottomrule
\end{tabular}
\caption{All tests significant, Delta Linear, Effect size 3}
\end{table*}

\begin{table*}
\centering
\centering
\begin{tabular}[t]{rrllllll}
\toprule
\multicolumn{1}{c}{ } & \multicolumn{1}{c}{ } & \multicolumn{2}{c}{Pmax\_BH} & \multicolumn{2}{c}{IntersectFDR} & \multicolumn{2}{c}{QCH} \\
\cmidrule(l{3pt}r{3pt}){3-4} \cmidrule(l{3pt}r{3pt}){5-6} \cmidrule(l{3pt}r{3pt}){7-8}
NbObs & Q & FDR & Power & FDR & Power & FDR & Power\\
\midrule
\rowcolor{gray!6}  1e+04 & 2 & 0.064 (0.016) & 0.346 (0.081) & 0.032 (0.01) & 0.232 (0.069) & 0.051 (0.008) & 0.349 (0.123)\\
1e+04 & 4 & 0 (0) & 0 (0) & 0.019 (0.028) & 0.018 (0.009) & 0.041 (0.013) & 0.166 (0.056)\\
\rowcolor{gray!6}  1e+04 & 8 & 0 (0) & 0 (0) & 0 (0) & 0 (0) & 0.004 (0.003) & 0.383 (0.107)\\
1e+05 & 2 & 0.066 (0.012) & 0.327 (0.074) & 0.033 (0.006) & 0.211 (0.063) & 0.05 (0.003) & 0.314 (0.105)\\
\rowcolor{gray!6}  1e+05 & 4 & 0 (0) & 0 (0) & 0.021 (0.015) & 0.018 (0.006) & 0.04 (0.006) & 0.166 (0.053)\\
1e+05 & 8 & 0 (0) & 0 (0) & 0 (0) & 0 (0) & 0.002 (0.001) & 0.259 (0.052)\\
\rowcolor{gray!6}  1e+06 & 2 & 0.064 (0.012) & 0.338 (0.074) & 0.033 (0.006) & 0.223 (0.066) & 0.05 (0.001) & 0.333 (0.115)\\
1e+06 & 4 & 0 (0) & 0 (0) & 0.02 (0.012) & 0.017 (0.006) & 0.039 (0.004) & 0.17 (0.053)\\
\rowcolor{gray!6}  1e+06 & 8 & 0 (0) & 0 (0) & 0 (0.002) & 0 (0) & 0.002 (0.001) & 0.238 (0.049)\\
\bottomrule
\end{tabular}
\caption{$Q-1$ among $Q$ significant tests, Delta Equal, Effect size 2}
\end{table*}

\begin{table*}
\centering
\centering
\begin{tabular}[t]{rrllllll}
\toprule
\multicolumn{1}{c}{ } & \multicolumn{1}{c}{ } & \multicolumn{2}{c}{Pmax\_BH} & \multicolumn{2}{c}{IntersectFDR} & \multicolumn{2}{c}{QCH} \\
\cmidrule(l{3pt}r{3pt}){3-4} \cmidrule(l{3pt}r{3pt}){5-6} \cmidrule(l{3pt}r{3pt}){7-8}
NbObs & Q & FDR & Power & FDR & Power & FDR & Power\\
\midrule
\rowcolor{gray!6}  1e+04 & 2 & 0.062 (0.015) & 0.82 (0.045) & 0.034 (0.008) & 0.751 (0.056) & 0.051 (0.004) & 0.807 (0.074)\\
1e+04 & 4 & 0.003 (0.005) & 0.096 (0.035) & 0.025 (0.015) & 0.479 (0.041) & 0.051 (0.007) & 0.716 (0.081)\\
\rowcolor{gray!6}  1e+04 & 8 & 0 (0) & 0 (0) & 0 (0) & 0.219 (0.024) & 0.039 (0.008) & 0.994 (0.012)\\
1e+05 & 2 & 0.061 (0.015) & 0.822 (0.049) & 0.033 (0.008) & 0.752 (0.063) & 0.05 (0.001) & 0.808 (0.081)\\
\rowcolor{gray!6}  1e+05 & 4 & 0.002 (0.002) & 0.096 (0.015) & 0.025 (0.014) & 0.481 (0.04) & 0.048 (0.003) & 0.707 (0.078)\\
1e+05 & 8 & 0 (0) & 0 (0) & 0 (0) & 0.221 (0.021) & 0.028 (0.004) & 0.992 (0.011)\\
\rowcolor{gray!6}  1e+06 & 2 & 0.062 (0.015) & 0.818 (0.049) & 0.034 (0.008) & 0.748 (0.062) & 0.05 (0.001) & 0.803 (0.08)\\
1e+06 & 4 & 0.002 (0.001) & 0.094 (0.01) & 0.024 (0.013) & 0.48 (0.039) & 0.048 (0.001) & 0.709 (0.079)\\
\rowcolor{gray!6}  1e+06 & 8 & 0 (0) & 0 (0) & 0 (0) & 0.219 (0.021) & 0.03 (0.004) & 0.993 (0.009)\\
\bottomrule
\end{tabular}
\caption{$Q-1$ among $Q$ significant tests, Delta Equal, Effect size 3}
\end{table*}


\begin{table*}
\centering
\centering
\begin{tabular}[t]{rrllllll}
\toprule
\multicolumn{1}{c}{ } & \multicolumn{1}{c}{ } & \multicolumn{2}{c}{Pmax\_BH} & \multicolumn{2}{c}{IntersectFDR} & \multicolumn{2}{c}{QCH} \\
\cmidrule(l{3pt}r{3pt}){3-4} \cmidrule(l{3pt}r{3pt}){5-6} \cmidrule(l{3pt}r{3pt}){7-8}
NbObs & Q & FDR & Power & FDR & Power & FDR & Power\\
\midrule
\rowcolor{gray!6}  1e+04 & 2 & 0.064 (0.015) & 0.899 (0.047) & 0.035 (0.008) & 0.855 (0.051) & 0.051 (0.004) & 0.895 (0.053)\\
1e+04 & 4 & 0.011 (0.006) & 0.753 (0.016) & 0.026 (0.013) & 0.82 (0.033) & 0.056 (0.009) & 0.953 (0.024)\\
\rowcolor{gray!6}  1e+04 & 8 & 0 (0) & 0.749 (0.014) & 0 (0) & 0.819 (0.028) & 0.053 (0.003) & 1 (0)\\
1e+05 & 2 & 0.064 (0.013) & 0.897 (0.041) & 0.036 (0.007) & 0.851 (0.046) & 0.05 (0.001) & 0.891 (0.05)\\
\rowcolor{gray!6}  1e+05 & 4 & 0.011 (0.005) & 0.758 (0.007) & 0.025 (0.014) & 0.825 (0.033) & 0.052 (0.003) & 0.956 (0.025)\\
1e+05 & 8 & 0 (0) & 0.748 (0.004) & 0 (0) & 0.813 (0.026) & 0.05 (0.002) & 1 (0)\\
\rowcolor{gray!6}  1e+06 & 2 & 0.062 (0.011) & 0.906 (0.036) & 0.034 (0.006) & 0.861 (0.04) & 0.05 (0) & 0.903 (0.04)\\
1e+06 & 4 & 0.012 (0.006) & 0.758 (0.008) & 0.026 (0.015) & 0.822 (0.034) & 0.052 (0.002) & 0.954 (0.025)\\
\rowcolor{gray!6}  1e+06 & 8 & 0 (0) & 0.748 (0.001) & 0 (0) & 0.811 (0.025) & 0.05 (0.001) & 1 (0)\\
\bottomrule
\end{tabular}
\caption{$Q-1$ among $Q$ significant tests, Delta Linear, Effect size 3}
\end{table*}


\end{document}